\documentclass[reqno,12pt]{article}
\usepackage{ae} 
\usepackage[T1]{fontenc}
\usepackage[ansinew]{inputenc}
\usepackage{amsmath}
\usepackage{amssymb}
\usepackage{graphicx}
\usepackage{color}
\usepackage{epsfig}
\usepackage{multirow}
\usepackage{pifont}

\newcommand{\beq}{\begin{equation}}
\newcommand{\eeq}{\end{equation}}
\newcommand\beqa{\begin{eqnarray}}
\newcommand\eeqa{\end{eqnarray}}
\newcommand\bea{\begin{array}}
\newcommand\eea{\end{array}}

\def\XXint#1#2#3{{\setbox0=\hbox{$#1{#2#3}{\int}$}
\vcenter{\hbox{$#2#3$}}\kern-.5\wd0}}

\newcommand{\neqa}{\nonumber\end{eqnarray}}

\renewcommand{\d}{\partial}

\newcommand{\<}{{\langle}}
\renewcommand{\>}{{\rangle}}

\newcommand{\re}{\relax{\rm I\kern-.18em R}}

\renewcommand{\sp}{p\hspace{-.40em}/}

\def\su2{{SU(2)}}

\def\[{\left[}
\def\]{\right]}

\def\({\left(}
\def\){\right)}
\def\[{\left[}
\def\]{\right]}

\def\<{\langle}
\def\>{\rangle}

\def\i2{\frac{i}{2}}

\newcommand{\ab}[1]{\langle #1\rangle}
\newcommand{\smallminus}{{\rm\rule[2.4pt]{6pt}{0.65pt}}}
\newcommand{\smallplus}{\hspace{0.5pt}\text{{\small+}}\hspace{-0.5pt}}
\renewcommand{\cap}{\mathrm{\raisebox{0.75pt}{{$\,\bigcap\,$}}}}

\def\spi{\relax{\rm \pi\kern-0.5em /}}
\def\sA{\relax{\rm A\kern-0.5em /}}
\def\sp{\relax{\rm p\kern-0.5em /}}
\def\sd{\relax{\rm \d\kern-0.5em /}}
\def\sk{\relax{\rm k\kern-0.5em /}}
\def\sn{\relax{\rm n\kern-0.5em /}}
\def\sl{\relax{\rm l\kern-0.5em /}}
\def\sP{\relax{\rm P\kern-0.7em /}}
\def\sBethe{\relax{\rm \Bethe\kern-0.5em /}}

\usepackage{varioref}
\usepackage{makeidx}
\makeindex

\usepackage[english]{babel}
\begin{document}


\thispagestyle{empty}

\renewcommand{\thefootnote}{\fnsymbol{footnote}}
\setcounter{footnote}{0} \setcounter{figure}{0}
\begin{center}
{\Large\textbf{\mathversion{bold} Some analytic results for two-loop scattering amplitudes }\par}

\vspace{1.0cm}

\textrm{Luis F. Alday}
\\ \vspace{1.2cm}
\footnotesize{

\textit{ School of Natural Sciences,\\Institute for Advanced Study, Princeton, NJ 08540, USA.} \\
\texttt{alday@ias.edu}}


\par\vspace{1.5cm}

\textbf{Abstract}\vspace{2mm}
\end{center}

\noindent

We present analytic results for the finite diagrams contributing to the two-loop eight-point MHV scattering amplitude of planar ${\cal N}=4$ SYM. We use a recently proposed representation for the integrand of the amplitude in terms of (momentum) twistors and focus on a restricted kinematics in which the answer depends only on two independent cross-ratios. The theory of motives can be used to vastly simplify the results, which can be expressed as simple combinations of classical polylogarithms.

\vspace*{\fill}

\setcounter{page}{1}
\renewcommand{\thefootnote}{\arabic{footnote}}
\setcounter{footnote}{0}

\newpage




\section{Introduction}

Over the last few years the study of scattering amplitudes of planar ${\cal N}=4$ SYM has received a lot of attention. Perhaps the most significant development regarding the perturbative side of the computation comes from applying the Grassmannian formalism in order to describe scattering amplitudes \cite{ArkaniHamed:2009si,ArkaniHamed:2009vw,Bullimore:2009cb}. Such efforts led to a proposal for an explicit recursive formula for the all loop integrand for scattering amplitudes in planar ${\cal N}=4$ SYM \cite{ArkaniHamed:2010kv} \footnote{See also \cite{Boels:2010nw}.}. Such developments are not only interesting per se, but may also help to make a connection with recent progress done in the strong coupling side of the computation \cite{Alday:2007hr,Alday:2010vh}.

Another interesting development regards the duality between MHV scattering amplitudes and the expectation value of certain Wilson loops. This was first observed at strong coupling \cite{Alday:2007hr} and then shown to hold also at one loop \cite{Drummond:2007aua,Brandhuber:2007yx}. The current status is that this duality has been tested at two loops for the case of six particles \cite{Drummond:2007bm,Bern:2008ap,Drummond:2008aq}. Due to the high complexity of the calculations, such test was done numerically. This duality triggered further research. In particular, in \cite{Anastasiou:2009kna}, an expression in terms of integrals for two-loop Wilson loops with any number of edges was given, making their numerical computation feasible. On the other hand, the analytic expression for the expectation value of the generic Wilson loop with six edges at two loops was obtained in \cite{DelDuca:2009au,DelDuca:2010zg}. The expression of \cite{DelDuca:2009au,DelDuca:2010zg} was given in terms of hundreds of Goncharov polylogarithms. Such polylogarithms satisfy non trivial identities, allowing, for instance, to write certain combinations of them in terms of classical polylogarithms. These identities can be efficiently taken into account by using the theory of motives \cite{motive}. In \cite{Goncharov:2010jf}, the theory of motives was used in order to write the expectation value of the six edged Wilson loop at two loops as a compact expressions, containing only classical polylogarithms.

Strong coupling computations \cite{Alday:2009yn} suggest that by focusing on a restricted kinematical regime one can obtain simpler answers. In this kinematical regime the momenta of the external particles live effectively on $1+1$ dimensions \footnote{Though the propagation of internal momenta is still four dimensional}. For the scattering of $n$ particles this leads to $n-6$ independent cross-ratios, as opposed to $3n-15$ for the generic case. Analytic expressions for the two-loop expectation value of Wilson loops in this kinematical regime were obtained, for the first non trivial case $n=8$ in \cite{DelDuca:2010zp} and then in full generality in \cite{Heslop:2010kq}.

Despite these developments, analytic results for two-loop scattering amplitudes are much harder to obtain. The idea of this note is very simple: we will consider the representation for two-loop amplitudes given in terms of momentum twistors integrals in \cite{ArkaniHamed:2010kv} and then focus in the simplest non trivial amplitude, namely, the eight-point, MHV amplitude in the restricted kinematical regime mentioned above. In this regime, the non trivial part of the amplitude depends on two cross-ratios. As we will see, diagrams of several topologies contribute to such amplitude: double-box, penta-box and double pentagons. Out of these, only the double pentagons are finite and we restrict our attention to those. Furthermore, for the restricted kinematics considered in this paper, only two types of double pentagon diagrams contribute. In this note we will present an analytic expression for such contribution. This should be the harder contribution\footnote{For instance, in the analog Wilson loop computation, the finite diagrams turn out to be also the harder to compute.}, though it doesn't need the introduction of a regulator, and we hope the other diagrams can be computed similarly.

This note is organized as follows. In section two we briefly review the representation for amplitudes as Grasmannian integrals given in \cite{ArkaniHamed:2010kv} and discuss some of the simplifications when considering the restricted kinematics. At the end of this section we describe the different contributions to the scattering of eight particles. 

In section three we proceed with the actual computation. We find that it is usually simpler to compute the action of a differential operator acting on the given contribution. After introducing Feynman parameters one can perform the integrals and express the answer as a sum of polylogarithms. Such expressions are usually very large. However, polylogarithms satisfy non trivial identities that can be efficiently taken into account by computing the symbol of a given expression. This leads to vast simplification and allows to write the final answer in a very compact analytic form. Finally, we end up with some conclusions. Many technical details, as well as a brief review on how to compute the symbol of relevant functions, are deferred to the appendices.

\section{Amplitudes as a Grasmannian integral and restricted kinematics}

In \cite{ArkaniHamed:2010kv} an expression for the integrand of all two-loop MHV amplitudes of planar MSYM was given. The expression is given in terms of the momentum twistors introduced in \cite{Hodges:2009hk}\footnote{Such variables are also natural from the strong coupling point of view \cite{Alday:2009yn}.} and has the following pictorial representation \footnote{I thank the authors of \cite{ArkaniHamed:2010kv} for the figure.}

\begin{eqnarray} \nonumber
\begin{split}
&\begin{array}{c@{~}c@{~}c@{~}c@{~}c@{~}c@{~}c@{~}c@{}}\raisebox{-1.25cm}{\includegraphics[scale=0.5]{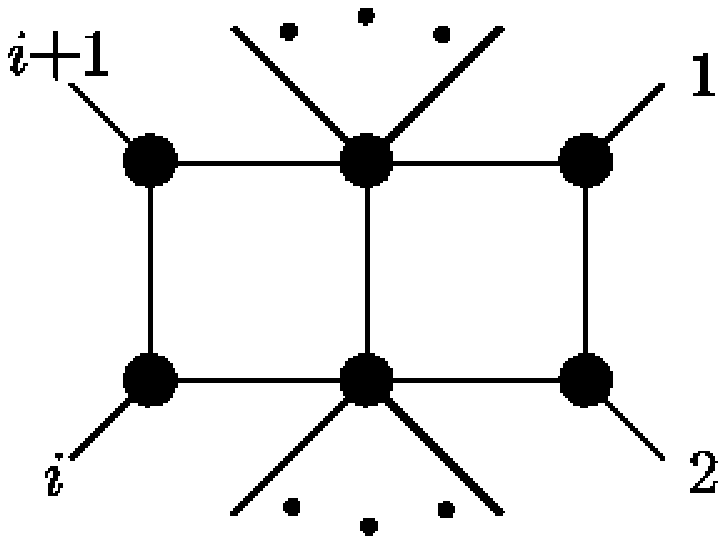}}&+&\hspace{-0.4cm}\raisebox{-1.5cm}{\includegraphics[scale=0.5]{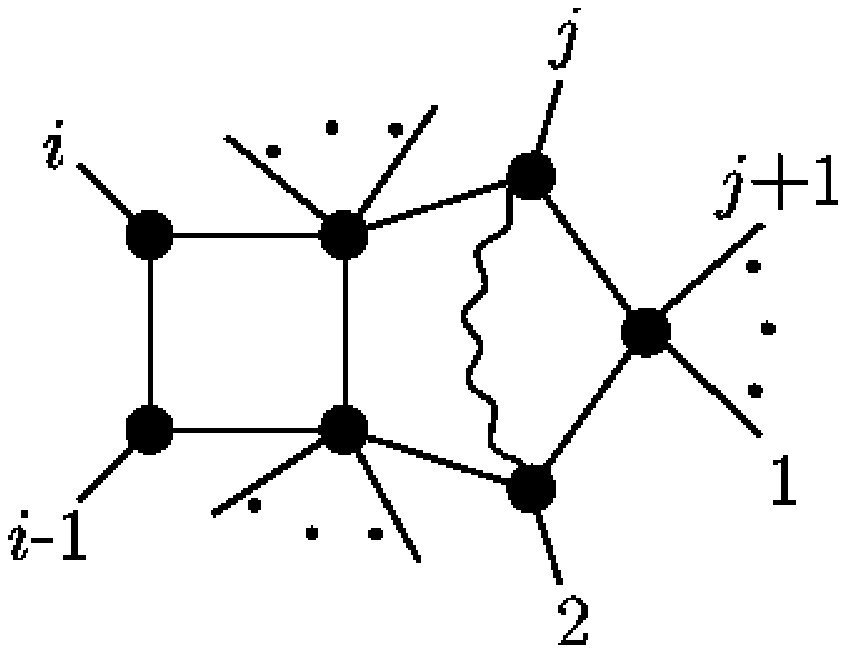}}&+&\raisebox{-1.5cm}{\includegraphics[scale=0.5]{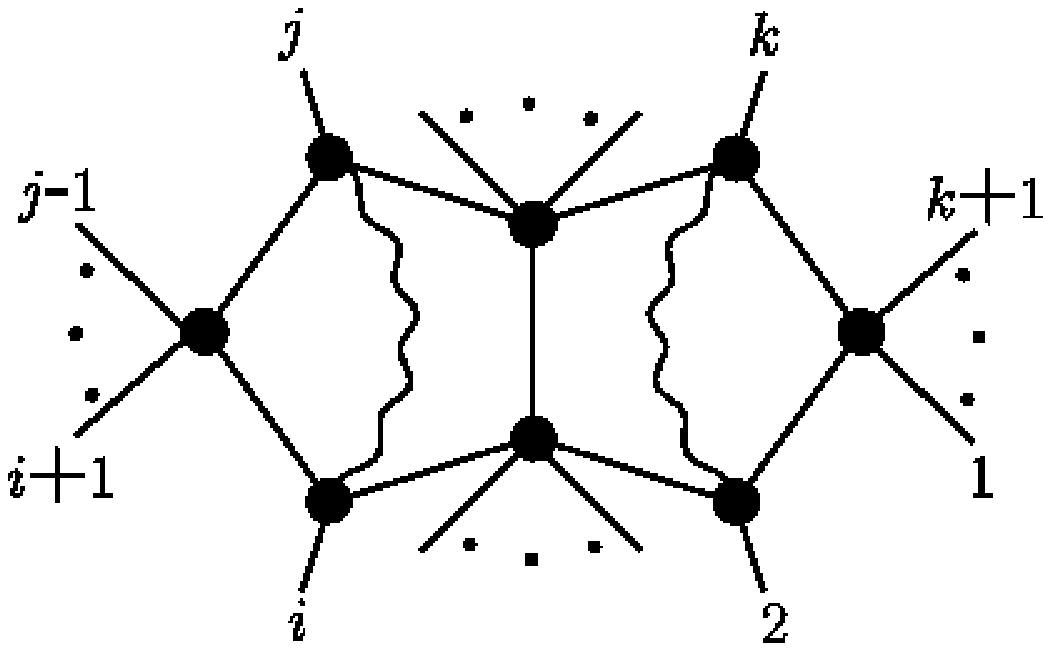}}\\\hspace{-0.5cm}
\begin{array}{c} \phantom{\times}\ab{n\,\,1\,\,2\,\,3} \times\\ \ab{1\,\,2\,\,i\,\,i\smallplus1} \ab{i\smallminus1\,\,i\,\,i\smallplus1\,\,i\smallplus2}\\~\\2<i<n\end{array}&&\hspace{-0.25cm}\begin{array}{c}\ab{2\,\,j\,\,i\smallminus1\,\,i} \ab{i\smallminus2\,\,i\smallminus1\,\,i\,\,i\smallplus1}\\\times \ab{AB|(123)\cap(j\smallminus1\,\,j\,\,j\smallplus1)}\phantom{\times}\\~\\3<i<j\leq n\end{array}&&\hspace{-0.0cm}\begin{array}{c}\ab{2\,\,i\,\,j\,\,k}\\\times \ab{AB|(123)\cap(k\smallminus1\,\,k\,\,k\smallplus1)}\phantom{\times}\\\times\ab{CD|(i\smallminus1\,\,i\,\,i\smallplus1)\cap(j\smallminus1\,\,j\,\,j\smallplus1)}\phantom{\times}\\2<i<j-1<k-1<n\end{array}\\\end{array}\end{split}
\end{eqnarray}
Each of these contributions is an integral over $dZ^{AB} dZ^{CD}$, where $(AB),(CD)$ run over lines in momentum twistor space \footnote{See appendix A for explicit computations.} and the following notation has been introduced
$$\langle AB|i-1 i i+1\cap k-1 k k+1\rangle=\langle A k-1 k k+1 \rangle \langle B i-1 i i+1 \rangle -(A \leftrightarrow B)$$
For each contribution the numerator of the integrand is written just below the figure, while the denominator is the standard denominator containing propagators and can be easily read from the figure. For instance, a generic double-box contribution looks like (for $i$ and $n-i$ sufficiently large)
\begin{equation}
\int \frac{dZ^{AB} dZ^{CD} \langle n 123 \rangle \langle 12i i+1 \rangle \langle i-1 i i+1 i+2 \rangle }{ \langle ABn1 \rangle \langle AB12 \rangle \langle AB23 \rangle \langle ABCD \rangle \langle CDi-1i \rangle \langle CDii+1 \rangle \langle CDi+1i+2 \rangle  }
\end{equation}
and so on.

\subsection{Restricted kinematics}

The above integrands are given in terms of momentum twistors.
We denote them by $\lambda^i$, with
$i=1,...,n$ and the brackets are defined as follows $$\langle ijkl\rangle=\epsilon^{\alpha \beta \gamma \delta} \lambda^i_\alpha\lambda^j_\beta\lambda^k_\gamma\lambda^l_\delta$$

The kinematics under consideration corresponds to the momenta of the external particles (or the location of the dual cusps)
lying in a $1+1$ plane. This can be achieved if we choose

\begin{equation}
\label{AdS3twistors}
\lambda^{2i}= \left(
                \begin{array}{c}
                  * \\
                  * \\
                  0 \\
                  0 \\
                \end{array}
              \right),~~~~~\lambda^{2i+1}=\left(
                \begin{array}{c}
                  0 \\
                  0 \\
                  * \\
                  * \\
                \end{array}
              \right)
\end{equation}

This special form implies certain simplifications. For instance, $\langle ijkl \rangle$
vanishes if three of the twistors involved have indices of
the same parity. On the other hand, four-brackets often factorize into products of two-brackets.
Another simplification is the following. Often in the
expressions for the amplitudes we find factors of the form $ \langle AB|
i-1,i,i+1 \cap j-1,j,j+1 \rangle$, where $A,B$ are generic (four dimensional) twistors
 and $i,j$ are of the form (\ref{AdS3twistors}). In the restricted kinematics we have the following remarkable identities

\begin{eqnarray}
\langle AB|
i-1,i,i+1 \cap j-1,j,j+1 \rangle &=& \langle ABij \rangle \langle i-1 i+1 j-1 j+1 \rangle,~~~i+j~ \mbox{odd} \nonumber \\
\langle AB|
i-1,i,i+1 \cap j-1,j,j+1 \rangle &=& \langle AB i-1 i+1 \rangle \langle i j-1 j j+1 \rangle \\ \nonumber
&=& \langle AB j-1 j+1 \rangle \langle i-1 i i+1 j  \rangle,~~~i+j~\mbox{even}
\end{eqnarray}
This implies a vast simplification for many of the numerators.

\subsection{Scattering of eight-particles}

In the restricted kinematics we can consider only the scattering
of an even number of particles. Given $n$ particles, we have
$n-6$ independent cross-ratios. Hence, the first non trivial case is the
scattering of eight particles. The non-trivial part of the scattering amplitude depends on two independent cross-ratios,
which we denote by $\chi^+$ and $\chi^-$.

The usual space-time cross-ratios $u_{ij}$ are given in terms of those by
\begin{eqnarray}
u_{15} &=&\frac{\chi^+}{1+\chi^+},~~~u_{26}=\frac{\chi^-}{1+\chi^-},~~~u_{37}=\frac{1}{1+\chi^+},~~~u_{48}=\frac{1}{1+\chi^-},~~~u_{i i+3}=1
\end{eqnarray}
where $u_{ij}$ can be written either in terms of dual coordinates or momentum twistors 
$$u_{ij}=\frac{x_{i,j+1}^2x_{i+1,j}^2}{x_{ij}^2 x_{i+1,j+1}^2}=\frac{\langle i,i+1,j+1,j+2 \rangle \langle i+1,i+2,j,j+1 \rangle}{ \langle i,i+1,j,j+1 \rangle \langle i+1,i+2,j+1,j+2 \rangle}$$

In appendix A we present a explicit choice for the momentum twistors.

We will focus in the double-pentagon diagram. Given the factor $\langle 2 i j k \rangle$ in the numerator, we see that the only non-vanishing contributions correspond to one of the indices $i,j,k$ being odd and the other two being even. This lead us to five contributions that can be separated into three kind of diagrams, see figure.

\begin{figure}[h]
\begin{center}
\includegraphics[width=140mm]{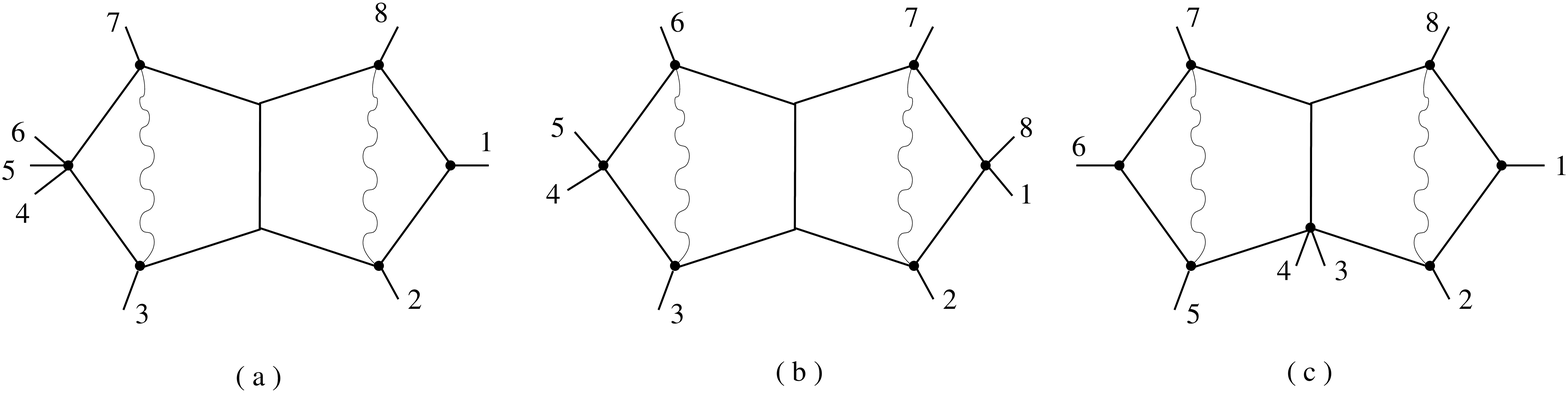}
\caption{Different kind of double pentagon diagrams contributing to the scattering of eight particles in the restricted kinematics. There are other two contributions that are the reflections of diagrams $(a)$ and $(c)$.}
\end{center}
\end{figure}

Diagrams of type $(a)$ are particulary simple for the case of the eight-point amplitude in restricted kinematics. Let us consider for instance the diagram plot in the figure, which corresponds to $i=3, j=7$ and $k=8$. We see that the twistor $\lambda_5$ does not appear in the integrand. We can always choose a gauge in which all the dependence on the cross-ratio $\chi^+$ is in $\lambda_5$, hence this contribution depends on $\chi^-$ only. Another way of saying the same, is that only three twistors appear with an odd label, and we need at least four in order to form a cross-ratio. As it will be clear in what follows, our computation is blind to this kind of terms. Furthermore we don't expect them in the final answer \footnote{For instance, if one believes the duality between Wilson loops and MHV scattering amplitudes holds for eight gluons at two loops, then such kind of terms should be absent.}, so they should cancel against similar contributions from other diagrams. It would be interesting to check this explicitly.

We will then focus on diagrams of type $(b)$ and $(c)$. There is only one diagram of type $(b)$ and corresponds to $i=3, j=6$ and $k=7$. Its contribution is given by
\begin{equation}
I^{(b)}=\int \frac{ dZ^{AB} dZ^{CD} \langle AB27 \rangle \langle CD36 \rangle \langle 2367 \rangle \langle 1368 \rangle \langle 2457 \rangle}{ \langle AB67 \rangle \langle AB78 \rangle \langle AB 12 \rangle \langle AB23 \rangle \langle ABCD \rangle \langle CD23 \rangle \langle CD34 \rangle \langle CD56 \rangle \langle CD67 \rangle}
\end{equation}
There are two diagrams of type $(c)$, related by a simple symmetry. For definiteness we focus in the one corresponding to $i=5,j=7$ and $k=8$, for which one obtains
\begin{equation}
\label{diagc}
I^{(c)}=\int \frac{ dZ^{AB} dZ^{CD} \langle AB13 \rangle \langle CD46 \rangle \langle 2781 \rangle \langle 5678 \rangle \langle 2578 \rangle}{ \langle AB78 \rangle \langle AB81 \rangle \langle AB 12 \rangle \langle AB23 \rangle \langle ABCD \rangle \langle CD45 \rangle \langle CD56 \rangle \langle CD67 \rangle \langle CD78 \rangle}
\end{equation}
In the next section we will focus on the computation of these two contributions.

\section{Double-pentagon contribution to the two-loop eight point amplitude in restricted kinematics}

\subsection{Diagram of type $(b)$}

As shown in appendix A, the computation of this diagram can be reduced to the computation of
\begin{eqnarray}
\label{doubpentsym}
& & I^{(z)}(\chi^+,\chi^-)=\int_0^\infty d\alpha_1...d\alpha_4 d \beta_1 ... d\beta_5 \beta_5 \frac{2\delta(\sum \alpha_i -1)\delta(\sum \beta_i -1)}{(\alpha_1+\alpha_2)(\alpha_3+\alpha_4)} \times \\ & & \times \frac{1}{ \left( ( \beta_2+(\alpha_1+\alpha_2)\beta_5+\beta_1 \chi^+ )(\beta_3+(\alpha_3+\alpha_4)\beta_5+\beta_4 \chi^-)+(\alpha_2+\alpha_3)(\beta_1+\beta_4)\beta_5 z\right)^3} \nonumber
\end{eqnarray}
As mentioned in the appendix, the sums inside the delta functions run over any subset of the $\alpha$'s and $\beta$'s, provided our choices keep the integral convergent. A judicious choice, however, can simplify the computation. Something that can also simplify the computation is to integrate the parameters in the right order. Often, one integration increases the transcendentality of the integrand by one. On the other hand, we have seven integration parameters and we know the final answer has transcendentality four, so three of the integrations can be done without increasing the transcendentality. Usually, it is preferable to push the increasing of transcendentality towards the end of the computation.

One should also keep in mind is that sometimes it is simpler to computed the action of differentials operators on the integrals we are interested in. For instance, for the case at hand we will see that it is much simpler to compute $\partial_{\chi^+} \partial_{\chi^-} I^{(z)}(\chi^+,\chi^-)$ rather than $I^{(z)}(\chi^+,\chi^-)$ \footnote{Of course, this means that our computation will be blind to functions that depend solely on one cross-ratio.}. This happens because at certain point in the computation, the action of certain differential operators (which depend only on the cross-ratios) lower the transcendentality of the integrand.

Coming back to the case at hand, note that the result is explicitly symmetric under interchange of $\chi^+$ and $\chi^-$. This is a consequence of a symmetry of the integrand under $\chi^+ \leftrightarrow \chi^-$ followed by $\beta_1 \leftrightarrow \beta_4$, $\beta_2 \leftrightarrow \beta_3$, $\alpha_1 \leftrightarrow \alpha_4 $ and $\alpha_2 \leftrightarrow \alpha_3$. We will choose which parameters add up to one and the order of integrations, in order to preserve this symmetry at every step of the computation.

Starting from (\ref{doubpentsym}), we introduce variables $\alpha^{\pm}=\alpha_1 \pm \alpha_2$ and $\tilde \alpha^{\pm}=\alpha_3 \pm \alpha_4$. Then we perform the integrations over $\alpha^-$ and $\tilde \alpha^-$ and use the delta function in order to set $\alpha^+ +\tilde \alpha^+=1$. Then, we integrate over $\beta_2$ and $\beta_3$ and use the delta function in order to set $\beta_5=1$. We arrive to the following expression

\begin{eqnarray}
& & I^{(z)}(\chi^+,\chi^-)=\int_0^\infty d \alpha^+ d \tilde \alpha^+ \delta(\alpha^+ + \tilde \alpha^+ - 1) d\beta_1 d \beta_4 \frac{1}{ z^2 \alpha^+ \tilde \alpha^+ (\beta_1+\beta_4)^2} \times \\ \nonumber
& & \times \left( Li_2\left(-\frac{R}{\alpha^+}\right)+Li_2 \left(-\frac{R}{\tilde\alpha^+}\right)-Li_2(-R)+\frac{1}{2}\log^2 R^+ R^--\log\left(z \alpha^+ \tilde\alpha^+ (\beta_1+\beta_4)\right)\log R^+ R^- \right. +\\
& & +\left. \frac{1}{2} \log^2(\beta_1+\beta_4)+\log(\beta_1+\beta_4)\log( z \alpha^+ \tilde \alpha^+)+\frac{1}{2}\left(\log^2 z+\log^2 \alpha^+ +\log^2 \tilde \alpha^+ \right)+\log z \log \alpha^+ \tilde \alpha^+ \right) \nonumber
\end{eqnarray}
where we have introduced $R^+=\alpha^+ + \beta_1 \chi^+,~R^-=\tilde \alpha^+ + \beta_4 \chi^-,~~  R=\frac{R^+ R^-}{z(\beta_1+\beta_4)}$ and we have dropped terms proportional to $\pi$. This can be done if one is interested in computing the highest functional transcendental part of the answer. As we will later see, the theory of motives is particularly suited for this purpose.

At this point we notice that acting with $\partial_{\chi^+} \partial_{\chi^-}$ lowers the transcendentality of the integrand by two! more precisely
\begin{eqnarray}
\partial_{\chi^+} \partial_{\chi^-} I^{(z)}(\chi^+,\chi^-)=\int_0^\infty d \alpha^+ d \tilde \alpha^+ \delta(\alpha^+ + \tilde \alpha^+ - 1) d\beta_1 d \beta_4 \left(\mbox{rational function} \right)
\end{eqnarray}
where the integral is some complicated but rational function. In order to compute $I^{(b)}$, we need to compute the above integral and its derivative w.r.t $z$ at $z=1$. It is convenient to compute $\partial_{\chi^+} \partial_{\chi^-} I^{(z)}(\chi^+,\chi^-)|_{z=1}$ and $\partial_z z^2 \partial_{\chi^+} \partial_{\chi^-} I^{(z)}(\chi^+,\chi^-)|_{z=1}$ and then consider the appropriate linear combination.

The final result appears to be quite large. It is expressed as the sum of several hundreds of terms, of the form $...+R_1 Li_2(R_2)+R_3 \log R_4 \log R_5+...$, where the $R's$ denote rational functions of $\chi^+$ and $\chi^-$. Polylogarithms satisfy many non trivial identities. As explained in \cite{Goncharov:2010jf,{motive}} and reviewed in appendix B, such identities can be taken efficiently into account by using the theory of motives and computing the symbol of the final answer. We obtain a remarkable simple result
\begin{eqnarray}
\label{Symbol}
S( \partial_{\chi^+} \partial_{\chi^-} I^{(b)})=\frac{1}{\chi^+ (\chi^+ - \chi^-)} \left(\chi^- \otimes (1+\chi^-) - \chi^- \otimes \chi^+ - (1+\chi^-) \otimes \chi^- \right. + \\+ \left. (1+\chi^-) \otimes \chi^+ - \chi^+ \otimes \chi^- + \chi^+ \otimes (1+\chi^-) \right) + (\chi^+ \leftrightarrow \chi^-) \nonumber
\end{eqnarray}
Knowing the symbol of a function suffices to determine the function up to terms that are functionally less transcendental, times constants of the appropriate transcendentality. By looking at the expressions one obtains, one can actually see that this is equivalent to disregard terms containing $\pi's$. \footnote{In principle, one is throwing terms like $\log 2$, etc. However, from the specific arguments that appear in the polygos, this doesn't seem to be the case.}

Given the symbol (\ref{Symbol}) we are then to guess a simple function with that symbol. The method is based on the symmetries of the symbol (see \cite{Goncharov:2010jf}). For the case at hand, we can decompose the symbol into its symmetric and antisymmetric components
\begin{equation}
a \otimes b =\frac{1}{2}( a \otimes b+ b \otimes a) + \frac{1}{2}( a \otimes b- b \otimes a)
\end{equation}
The function we are looking for has transcendentality two, so it is formed by a sum of terms of the form $Li_2(z)$ and $\log(z)\log(y)$. From the definition of the symbol it is clear that only the terms of the form $Li_2(z)$ contribute to the antisymmetric component of the symbol. The antisymmetric component of the symbol (\ref{Symbol}) is simply
\begin{equation}
S( \partial_{\chi^+} \partial_{\chi^-} I^{(b)})^{anti}= \frac{1}{\chi^+ (\chi^+ - \chi^-)} \left( \chi^- \otimes (1+\chi^-) - (1+\chi^-) \otimes \chi^-  \right)+(\chi^+ \leftrightarrow \chi^-)
\end{equation}
One can immediately guess a simple function having that symbol, namely $\frac{2}{\chi^+(\chi^--\chi^+)}Li_2(-\chi^-)$ symmetrized under $\chi^+ \leftrightarrow \chi^-$. Now we can subtract this to the total answer and find the combination of logarithms leading to the symmetric part of the new symbol. The final answer is remarkably simple
\begin{equation}
\partial_{\chi^+} \partial_{\chi^-} I^{(b)}= \frac{1}{\chi^+ (\chi^+ - \chi^-)} \left(2 Li_2(-\chi^-) +\log(1+\chi^-) \log \chi^+ \chi^- -\log \chi^+ \log \chi^- \right)+(\chi^+ \leftrightarrow \chi^-)
\end{equation}
Several comments are in order. First, it is interesting that a single diagram, even before summing the other contributions, is relatively simple. Second, note the somewhat unpleasant feature that the answer is not a sum of factorized terms, of the form $f(\chi^+)g(\chi^-)$, due to the presence of $(\chi^+-\chi^-)$ in the denominator. In particular this makes difficult to extract $I^{(b)}$ from its double derivative. On the other hand, one can integrate one of the two variables, lets say $\chi^-$ and check explicitly that the answer is of the form $\frac{1}{\chi^+} (Transcendental~functions)$. This in particular implies that the final answer will have transcendentality four, with no rational functions in front of the transcendental functions. This is guaranteed by the construction of \cite{ArkaniHamed:2010kv}, but for the point of view of the explicit computation looks rather non trivial.

\subsection{Diagram of type $(c)$}

We now turn our attention to the computation of diagram (c), eq. (\ref{diagc}). The computation proceeds along the same lines as the previous computation. We found it convenient to choose a gauge where the dependence on the cross-ratios is on $\lambda_4$ and $\lambda_5$. Along the lines of the computation of appendix A, it is immediate to write an expression for $I^{(c)}$ in terms of Feynman parameter, though it is pretty lengthy to be presented here. We find that it is simpler to compute $\partial_{\chi^+}I^{(c)}$. As expected, the final answer has transcendentality three. Again, it is pretty lengthy, but its symbol is much simpler:
{\footnotesize
\begin{eqnarray*}
& & S(\chi^+ (1+\chi^+)\partial_{\chi^+}I^{(c)} )= \\
& & -\chi^- \otimes (1+\chi^-)  \otimes (1- \chi^+ \chi^-) -\chi^- \otimes \chi^+ \otimes (1- \chi^+ \chi^-) -\chi^- \otimes (1+\chi^+) \otimes (1+\chi^-)+\\
 & & +\chi^- \otimes (1+\chi^+) \otimes (1- \chi^+ \chi^-)  +(1+\chi^-)  \otimes \chi^- \otimes (1- \chi^+ \chi^-)+(1+\chi^-)  \otimes \chi^+ \otimes (1- \chi^+ \chi^-)-\\
   & & -(1+\chi^-)  \otimes (1+\chi^+) \otimes \chi^+ -\chi^+ \otimes \chi^- \otimes (1- \chi^+ \chi^-)  +\chi^+ \otimes (1+\chi^-)  \otimes (1- \chi^+ \chi^-)+ \\
   & & + \chi^+ \otimes (1+\chi^+) \otimes (1+\chi^-) -\chi^+ \otimes (1+\chi^+) \otimes (1- \chi^+ \chi^-) -(1+\chi^+) \otimes \chi^- \otimes (1+\chi^-)+\\
     & & +(1+\chi^+) \otimes \chi^- \otimes (1- \chi^+ \chi^-)-(1+\chi^+) \otimes (1+\chi^-)  \otimes \chi^+ -(1+\chi^+) \otimes \chi^+ \otimes (1+\chi^-) +\\
      & & +(1+\chi^+) \otimes \chi^+ \otimes (1- \chi^+ \chi^-)
\end{eqnarray*}
}
As before, a function with the correct symbol can be guessed by making use of the symmetries. Denoting different components by $i \otimes j \otimes k$ the following table shows which functions contribute to different combinations. $S_{ijk}$ denote the completely symmetric combination, $A_{ij}$ denotes the antisymmetric component in the first two indices and $A_{jk}$ the antisymmetric component in the last two.
\begin{center}
 \begin{tabular}{| l | c | c | r| }
  \hline  & $S_{ijk}$ & $A_{ij}$ & $A_{ik}$ \\
   \hline $Li_3(x)$ & $\checkmark$ & $\checkmark$ & - \\
   \hline $Li_2(x)\log y $ & $\checkmark$ & $\checkmark$  & $\checkmark$ \\
  \hline $\log x \log y \log z$ & $\checkmark$ & - & - \\
   \hline \end{tabular}
   \end{center}
The strategy is quite clear. We first compute the component of the symbol that is antisymmetric in the last two indices and try to guess a function of the form $\sum Li_2(z) \log x$ which takes care of that component. Once we have subtracted this contribution we can distinguish between $\log x \log y \log z$ and $Li_3(x)$ by their symmetry under interchange of the first two components. Following this procedure we arrive at our final result, which we present in appendix C.  It is quite lengthy but it has all the correct features. In particular, when integrating $\chi^+$, one obtains functions of the right transcendentality, without multiplicative rational functions.

For the analysis of the next subsection, we will need $\partial_{\chi^+} \partial_{\chi^-} I^{(c)}$, which is given by
\begin{eqnarray*}
& & \chi^+ (1+\chi^+ )(1+\chi^-) (\chi^+ \chi^- -1)\partial_{\chi^+} \partial_{\chi^-} I^{(c)} =\\
& & -\chi^- \chi^+ \log (\chi^-) \log (\chi^-+1)-\chi^+ \log (\chi^-) \log (\chi^-+1)+\chi^- \chi^+ \log (\chi^+) \log (\chi^-+1)+ \\
& & +\chi^+ \log (\chi^+) \log (\chi^-+1)-\chi^- \chi^+ \log (\chi^-) \log (\chi^+)-\chi^+ \log (\chi^-) \log (\chi^+)+\chi^+ \log (\chi^-) \log (\chi^++1)+\\
& & +\log (\chi^-) \log (\chi^++1)-\chi^+ \log
   (\chi^+) \log (\chi^++1)-\log (\chi^+) \log (\chi^++1)-\\
   & & -2 \left((\chi^-+1) \chi^+ \text{Li}_2(-\chi^-)+(\chi^++1) \text{Li}_2(-\chi^+)\right)
   \end{eqnarray*}
Again, note that the factor $(\chi^+ \chi^- -1)$ in the denominator prevents us from writing the answer in a factorized form.

\subsection{Combining everything}

We now combine all diagrams into the total answer. For that we need to sum over cyclic permutations, counting only once diagrams with identical integrands \footnote{After symmetrizing the variables of integration $Z^{AB} \leftrightarrow Z^{CD}$.}.

Once a specific diagram is computed, cyclic permutations are very easy to take into account, as each diagram is a function of the cross-ratios only. Cyclic permutations simply correspond to simple transformations among the cross-ratios. More specifically, if the result of a diagram is $I(\chi^+,\chi^-)$, then, the total sum over cyclic permutations is
\begin{equation}
\sum_{cyclic} I = I(\chi^+,\chi^-)+ I(\chi^- , \frac{1}{\chi^+})+ I (\frac{1}{\chi^+},\frac{1}{\chi^-})+I(\frac{1}{\chi^-},\chi^+)
\end{equation}
In this paper we have computed quantities of the form $\partial_{\chi^+} \partial_{\chi^-} I(\chi^+,\chi^-) \equiv I_{+-}(\chi^+,\chi^-) $. When adding them up, we should take into account the appropriate prefactors from taking the derivatives. Our final result is
\begin{eqnarray} \nonumber
\partial_{\chi^+} \partial_{\chi^-}I^{total} = I_{+-}(\chi^+,\chi^-) -\frac{I_{+-}(1/\chi^-,\chi^+)}{(\chi^-)^2} -\frac{I_{+-}(\chi^-,1/\chi^+)}{(\chi^+)^2}+\frac{I_{+-}(1/\chi^+,1/\chi^-)}{(\chi^+ \chi^-)^2}
\end{eqnarray}
with
\begin{equation}
 I_{+-}(\chi^+,\chi^-)= \partial_{\chi^+} \partial_{\chi^-} I^{(b)}+ 2 \partial_{\chi^+} \partial_{\chi^-} I^{(c)}
\end{equation}
The factor of 2 in front of $I^{(c)}$ is due to the fact that we have two diagrams of such type. Using the explicit expressions we obtain
\begin{eqnarray}
\label{final}
& & \partial_{\chi^+} \partial_{\chi^-}I^{total} = \\
& & \frac{1}{\chi^+ \chi^-} \log \chi^+ \log \chi^- + \frac{2(\chi^+ -1)}{\chi^+ \chi^-(1+\chi^+)} \log \chi^- \log(1+\chi^-)+ \frac{2(\chi^- -1)}{\chi^+ \chi^- (1+\chi^-)} \log \chi^+ \log(1+\chi^+) \nonumber +\\
& & + \frac{4(\chi^+-1)}{\chi^+ \chi^-(1+\chi^+)}\text{Li}_2(-\chi^-)+ \frac{4(\chi^- -1)}{\chi^+ \chi^- (1+\chi^-)} \text{Li}_2(-\chi^+) \nonumber
\end{eqnarray}
Something remarkable has happened. Note that terms like $(\chi^+ -\chi^-)$ or $(1 - \chi^+ \chi^-)$ have disappeared from the denominator and each term is of the factorized form $f(\chi^+)g(\chi^-)$! in particular, it is immediate to integrate on $\chi^+$ and $\chi^-$. The final result is expressed as a sum of terms of transcendentality four, each of them of the factorized form $f(\chi^+)g(\chi^-)$, and with no rational functions in front of them. As an aside remark, we note that the answer does not contain tetra-logarithms.\footnote{If the duality between Wilson loops and scattering amplitudes is to hold for this case, then one should have canceled these tetra-logarithms from contributions coming from somewhere else. However, we expect the other, somehow simpler, diagrams, to contain only less transcendental functions.}

\section{Conclusions}

We considered the two-loop eight-point MHV scattering amplitude of planar ${\cal N}=4$ SYM. We used the representation of \cite{ArkaniHamed:2010kv}, which gives such amplitude as certain integral over twistorial variables. The amplitude receives three kind of contributions, coming from double-boxes, penta-boxes and double-pentagon diagrams. The only finite contribution is the one corresponding to double-pentagons. On the other hand, at least naively, these diagrams seem to be the harder or more intricate, and we focus on those. There exist a convenient kinematical restriction, in which each contribution is a function of two independent cross-ratios only. In this note we have obtained analytical results for the double-pentagon contributions in such restricted kinematics, see (\ref{final}). Apart from results in certain kinematical limits, it seems very hard to find explicit analytical results for scattering amplitudes at two loops \footnote{See however \cite{Drummond:2010mb}.}.

We can draw some conclusions from our results. First, it is remarkable that the answers, even for single diagrams, are quite simple. This suggest that the kinematical restriction considered in this paper may be a very good arena in order to study the representation given in \cite{ArkaniHamed:2010kv}. Second, along our computation we have seen that it is usually convenient to study, rather than the contributions themselves, the result of certain differential operators acting on those contributions. This is of course related to the fact that these contributions are expected to satisfy certain differential equations, as mentioned in \cite{ArkaniHamed:2010kv}, due to their Yangian invariance. Third, we have seen that the theory of motives can be efficiently used in order to simplify the computation at various stages. It would be very interesting to learn how to compute the symbol of a given contribution directly, without having to compute the relevant integrals. Related to this, it is suggestive that identities among polylogarithms (at least for the dilogarithmical case) are known to be related to the TBA for certain integrable systems, see for instance \cite{Gliozzi:1994cs}. It would be fascinating if this helps to make a connection with similar structures arising at strong coupling \cite{Alday:2009dv}.

One of the motivations to study this problem was to check the duality between Wilson loops and MHV scattering amplitudes for this case. The answer for the expectation value of the eight-sided Wilson loop is known in this restricted kinematics \cite{DelDuca:2010zp} as is quite simple. Once all diagrams are added up, one should reproduce this result from the scattering amplitude computation, if the duality holds. In this note we have computed the harder diagrams, in the sense that they require the highest amount of Feynman parameters \footnote{In the analog Wilson loop computations, the finite diagrams were also the harder to compute}. The computation of the other diagrams could in principle be done along similar lines to the ones described in these notes, though they require the introduction of a regulator, since they are divergent.  As suggested by explicit computations, see for instance \cite{Mason:2010pg,Drummond:2010mb}, a convenient regularization is to go to the Coulomb branch \cite{Alday:2009zm}.  These diagrams are currently being computed and will be presented in a forthcoming publication.

The result found in this note, even if partial, shares a very interesting feature with the Wilson loop answer. Namely, the result is of the factorized form $\sum f(\chi^+) g(\chi^-)$. This form is not at all obvious and it actually doesn't hold before we add up all double pentagon diagrams! it would be very interesting to find a representation where this factorization is manifest.

Besides the open problems already mentioned, one would be to extend the computation done here to the generic case with restricted kinematics. Another interesting problem would be to understand how the constraints from the OPE for Wilson loops \cite{Alday:2010ku} manifest themselves on the integrals of \cite{ArkaniHamed:2010kv}. The requirement of unit leading singularity for each integral, plus the constraints from the OPE (which are probably related!) would presumably restrict which kind of functions can appear in the answer and the arguments of these functions. For instance, in the present computation, we experimentally observe that the  symbol of the answer is always build up from a few building blocks, like $\chi^+$ or $(1+\chi^+)$, but never $(2+\chi^+)$.

\section*{Acknowledgments}
I would like to thank J. Bourjaily for collaborations at early stages of this project. The author has also benefited from discussions with  N. Arkani-Hamed, S. Caron-Huot and J. Maldacena. The author is supported in part by the DOE grant DE-FG02-
90ER40542.


\appendix

\section{Some details on the explicit computations}

\subsection{One loop example}

In this appendix we show some explicit details on the computation of the relevant integrals. Let us start, as a warmup, with the following one loop pentagon integral, see figure

\begin{figure}[h]
\begin{center}
\includegraphics[width=40mm]{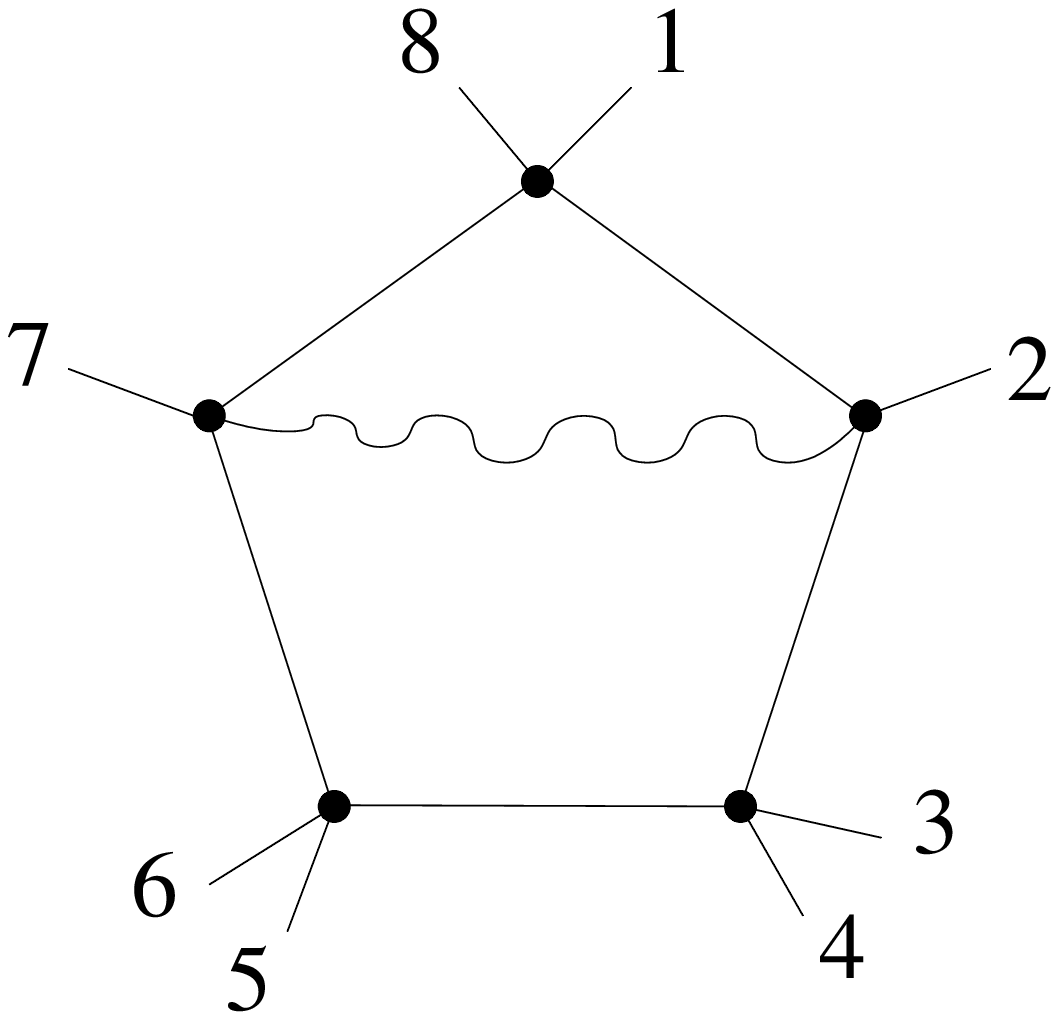}
\end{center}
\end{figure}
Its contribution is
\begin{equation}
\label{penta}
I_{p}^{(1)}= \int \frac{dZ^{AB} \langle AB27 \rangle \langle 2745 \rangle \langle 1368 \rangle}{ \langle AB12 \rangle \langle AB23 \rangle \langle AB45 \rangle \langle AB67 \rangle \langle AB78 \rangle}
\end{equation}
We will make use of the Feynman parametrization

\begin{equation}
\frac{1}{A_1^{\lambda_1} ... A_n^{\lambda_n}}=\frac{\Gamma(\lambda_1+...+\lambda_n)}{\Gamma(\lambda_1)...\Gamma(\lambda_n)} \int_0^\infty d\alpha_1 ...d\alpha_n \delta(\sum \alpha_i -1) \frac{\alpha_1^{\lambda_1 -1}...\alpha_n^{\lambda_n -1}}{(\alpha_1 A_1+...+\alpha_n A_n)^{\lambda_1+...+\lambda_n}}
\end{equation}
where the delta function includes any subset of $\alpha's$. As far as the integral is convergent, the result does not depend on such a choice. Using the Feynman parametrization we can convert (\ref{penta}) into
\begin{eqnarray}
I_{p}^{(1)}=\frac{\langle 2745 \rangle \langle 1368 \rangle}{\langle 1245 \rangle \langle 2345 \rangle\langle 6745 \rangle \langle 7845 \rangle} \Gamma(5) \int dZ^{AB} d \alpha_1 ... d \alpha_5 \delta(\sum \alpha_i -1) \frac{ \langle AB27 \rangle}{ \langle AB Y \rangle^5}\\
Y= \alpha_1 \frac{X_{12}}{\langle 1245 \rangle}+\alpha_2 \frac{X_{23}}{\langle 2345 \rangle}+\alpha_3 \frac{X_{67}}{\langle 6745 \rangle}+\alpha_4 \frac{X_{78}}{\langle 7845 \rangle}+\alpha_5 X_{45} \nonumber
\end{eqnarray}
now we are on position of performing to integral over $Z^{AB}$, see for instance \cite{Hodges:2010kq,Mason:2010pg,{Drummond:2010mb}}.
\begin{equation}
\int dZ^{AB} \frac{ \langle AB27 \rangle}{ \langle AB Y \rangle^5} =-\frac{1}{4} X_{27}\partial_Y \int dZ^{AB} \frac{1}{ \langle AB Y \rangle^4}=-\frac{1}{6} X_{27}\partial_Y \frac{1}{ \langle YY \rangle^2} =\frac{2}{3} \frac{ \langle 27 Y\rangle}{\langle YY \rangle^3}
\end{equation}
At this point we are left with the integrals over the Feynman parameters. One can easily see that the integral over $\alpha_5$ can be done immediately and it is convenient to set $\sum_{i=1}^4 \alpha_i=1$. We obtain
\begin{equation}
I_{p}^{(1)}=  2\frac{\langle 2745 \rangle^2 \langle 1368 \rangle}{\langle 1245 \rangle \langle 2345 \rangle\langle 6745 \rangle \langle 7845 \rangle} \int d \alpha_1 ... d \alpha_4 \delta(\sum \alpha_i -1) \frac{1}{\langle \tilde Y \tilde Y \rangle}
\end{equation}
with $\tilde Y=Y(\alpha_5=0)$. The integral will be a function of the cross-ratios and it should not depend on our specific choice for the momentum twistors. However, a judicious choice may simplify the computation. We choose
\begin{eqnarray}
\lambda_1=\left(
                \begin{array}{c}
                  0 \\
                  0 \\
                  i \sqrt{2}\chi^+ \\
                  \frac{i}{\sqrt{2}}(1-\chi^+) \\
                \end{array}
              \right),~~~
\lambda_2=\left(
                \begin{array}{c}
                  0 \\
                  \frac{i}{\sqrt{2}} \\
                  0 \\
                  0 \\
                \end{array}
              \right),~~~
              \lambda_3=\left(
                \begin{array}{c}
                  0 \\
                  0 \\
                  -i \sqrt{2} \\
                  \frac{i}{\sqrt{2}} \\
                \end{array}
              \right),~~~
              \lambda_4=\left(
                \begin{array}{c}
                  i \sqrt{2} \\
                  -i \sqrt{2} \\
                  0 \\
                  0 \\
                \end{array}
              \right)\\
              \lambda_5=\left(
                \begin{array}{c}
                  0 \\
                  0 \\
                  i \sqrt{2} \\
                  -i \sqrt{2} \\
                \end{array}
              \right),~~~
              \lambda_6=\left(
                \begin{array}{c}
                  -i \sqrt{2} \\
                  \frac{i}{\sqrt{2}} \\
                  0 \\
                  0 \\
                \end{array}
              \right),~~~
                  \lambda_7=\left(
                \begin{array}{c}
                  0 \\
                  0 \\
                  0 \\
                  \frac{i}{\sqrt{2}} \\
                \end{array}
              \right),~~~
                  \lambda_8=\left(
                \begin{array}{c}
                  i \sqrt{2} \chi^- \\
                  \frac{i}{\sqrt{2}}(1-\chi^-) \\
                  0 \\
                  0 \\
                \end{array}
              \right) \nonumber
\end{eqnarray}
The space-time cross-ratios $u_{ij}=\frac{x_{i,j+1}^2x_{i+1,j}^2}{x_{ij}^2 x_{i+1,j+1}^2}=\frac{\langle i,i+1,j+1,j+2 \rangle \langle i+1,i+2,j,j+1 \rangle}{ \langle i,i+1,j,j+1 \rangle \langle i+1,i+2,j+1,j+2 \rangle}$ are given by
\begin{equation}
u_{15} =\frac{\chi^+}{1+\chi^+},~~~u_{26}=\frac{\chi^-}{1+\chi^-},~~~u_{37}=\frac{1}{1+\chi^+},~~~u_{48}=\frac{1}{1+\chi^-},~~~u_{i i+3}=1
\end{equation}
returning to our integral, with the above choice we obtain
\begin{equation}
I_{p}^{(1)}= -\frac{(1+\chi^+)(1+\chi^-)}{1+\chi^+ + \chi^- +\chi^+ \chi^-} \int d \alpha_1 ... d \alpha_4  \frac{\delta(\sum \alpha_i -1)}{(\alpha_2+\chi^+ (\alpha_1+\alpha_2))(\alpha_2+\chi^-(\alpha_3+\alpha_4))}
\end{equation}
The integral is manifestly symmetric under interchange of $\chi^+$ and $\chi^-$. By using mathematica we can readily perform it and we obtain
\begin{equation}
I_{p}^{(1)}=-\log\left(1+\frac{1}{\chi^+} \right)\log\left(1+\frac{1}{\chi^-} \right)
\end{equation}
As expected, we obtain a result of transcendentality two. Furthermore, note that there is no rational function in front of the transcendental functions. This remarkable feature is true (by construction ) for each integral in the amplitude representation of \cite{ArkaniHamed:2010kv}, which we are using.

\subsection{Two-loop double-pentagon example}

We want to compute the following two-loop double pentagon integral, which was called $I^{b}$ in the main text.

\begin{equation}
I_{pp}^{(2)}=\int \frac{ dZ^{AB} dZ^{CD} \langle AB27 \rangle \langle CD36 \rangle \langle 2367 \rangle \langle 1368 \rangle \langle 2457 \rangle}{ \langle AB67 \rangle \langle AB78 \rangle \langle AB 12 \rangle \langle AB23 \rangle \langle ABCD \rangle \langle CD23 \rangle \langle CD34 \rangle \langle CD56 \rangle \langle CD67 \rangle}
\end{equation}
Introducing Feynman parameters as before we obtain
\begin{equation}
I_{pp}^{(2)}=32 {\cal N} \int d\alpha_1...d\alpha_4 d \beta_1 ... d\beta_5 \beta_5 \frac{\delta(\sum \alpha_i -1)\delta(\sum \beta_i -1)}{\langle \tilde Y \tilde Y \rangle}\left(6\frac{\langle 27Q \rangle \langle 36Q \rangle}{\langle QQ \rangle^4}-\frac{\langle 2736 \rangle}{\langle QQ \rangle^3} \right)
\end{equation}
where ${\cal N}=1$ for our choice of twistors and we have introduced the following components
\begin{eqnarray}
\tilde Y &=& \alpha_1 X_{23}+\alpha_2 X_{34}+\alpha_3 X_{56}+\alpha_4 X_{67},\\
Q&=&\beta_1 X_{12}+\beta_2 X_{23}+\beta_3 X_{67}+\beta_4 X_{78}+\beta_5 \tilde{Y}
\end{eqnarray}
The simplicity of the restricted kinematics we are considering becomes apparent when specifying the above expressions for our choice of gauge
\begin{eqnarray}
\langle \tilde Y \tilde Y \rangle = 2(\alpha_1+\alpha_2)(\alpha_3+\alpha_4),~~~~~\langle 27Q \rangle \langle 36Q \rangle= (\alpha_2+\alpha_3)(\beta_1+\beta_4)\beta_5 \\
\langle QQ \rangle = 2 ( \beta_2+(\alpha_1+\alpha_2)\beta_5+\beta_1 \chi^+ )(\beta_3+(\alpha_3+\alpha_4)\beta_5+\beta_4 \chi^-)+2(\alpha_2+\alpha_3)(\beta_1+\beta_4)\beta_5
\end{eqnarray}
It is then clear that the integral is symmetric under the interchange of $\chi^+$ and $\chi^-$. As a technical trick, we introduce a new variable $z$ and consider
\begin{eqnarray}
& & I^{(z)}=\int d\alpha_1...d\alpha_4 d \beta_1 ... d\beta_5 \beta_5 \frac{2\delta(\sum \alpha_i -1)\delta(\sum \beta_i -1)}{(\alpha_1+\alpha_2)(\alpha_3+\alpha_4)} \times \\ \nonumber & &\times \frac{1}{ \left( ( \beta_2+(\alpha_1+\alpha_2)\beta_5+\beta_1 \chi^+ )(\beta_3+(\alpha_3+\alpha_4)\beta_5+\beta_4 \chi^-)+(\alpha_2+\alpha_3)(\beta_1+\beta_4)\beta_5 z\right)^3}
\end{eqnarray}
Then $I_{pp}^{(2)}=-\left(\partial_z I^{(z)}+I^{(z)}\right)|_{z=1}$,as used in the main text.

\section{Symbol of transcendental functions}

Scattering amplitudes of planar MSYM at a given loop order are believed to be functions of the cross-ratios of homogeneous transcendentality. All known perturbative results can be expressed in terms of polylogarithms of rational functions of the cross-ratios. A polylogarithm $Li_k(z)$ is said to have transcendentality $k$, while the transcendentality of a product of two functions is the sum of their transcendentalities. Polylogarithms satisfy non trivial identities. In order to study such identities, the concept of symbol is very powerful. The symbol of a function of transcendentality $k$ is an element of the $k-$fold tensor product of the multiplicative group of rational functions modulo constants \cite{motive}. The symbol of a polylogarithms is
\begin{equation}
{\rm S}( Li_k(z)) = - (1 - z) \otimes
\underbrace{z \otimes \cdots \otimes z}_{k-1~{\rm times}}.
\end{equation}
while ${\rm S}(\log(z))=z$, as $\log(z)=-Li_1(1-z)$. The symbol satisfies a shuffle algebra, so that, for instance
\begin{eqnarray}
{\rm S}(\log(z_1) Li_k(z_2))&=& -z_1 \otimes (1-z_2) \otimes z_2 \otimes ... \otimes z_2- \\ & &- (1-z_2) \otimes z_1 \otimes z_2 \otimes...\otimes z_2-...-(1-z_2)\otimes z_2 \otimes...\otimes z_2 \otimes z_1 \nonumber
\end{eqnarray}
so the symbol of $\log(z_1)$ is inserted in each possible way. In the body of the paper we consider transcendental functions whose arguments are rational functions of the cross-ratios which we will generically denote as $z_i$. The symbol satisfies
\begin{eqnarray}
A \otimes \frac{z_i z_j}{z_k} \otimes B = A \otimes z_i \otimes B + A \otimes z_j \otimes B - A \otimes z_k \otimes B \\
A \otimes c z \otimes B = A \otimes z \otimes B
\end{eqnarray}
where $A,B$ can be composed symbols and $c$ is a constant. So, for instance, the symbols of $\log (z)$ , $\log (2 z)$ and $\log(-z)$ all coincide. The knowledge of the symbol of a function does not fix uniquely the function, it fixes it only up to functions of lower transcendentality times constants of the appropriate transcendentality.

As already mentioned, the symbol is a very powerful tool for studying the identities among polylogarithms. For instance, lets compute the symbol of the following function
\begin{eqnarray}
{\rm S}( Li_2(z) + Li_2(1-z))=-(1-z)\otimes z - z \otimes (1-z)
\end{eqnarray}
This can be seen to be equal to the symbol of $-\log (z) \log (1-z)$. Indeed, a well known identity of dilogarithms says that the two functions differ by a term proportional to $\pi^2$, which is invisible to the symbol.

\section{Result for $\partial_{\chi^+}I^{(c)}$}

{\small
\begin{eqnarray*}
& & \chi^+(1+\chi^+)\partial_{\chi^+}I^{(c)} = \\
& & \frac{1}{6} \log ^3(\chi^-+1)-\frac{1}{2} \log (\chi^-) \log ^2(\chi^-+1)-\frac{1}{2} \log (\chi^++1) \log ^2(\chi^-+1)-\frac{1}{2} \log ^2(\chi^++1) \log (\chi^-+1)-\\
& & - \log ^2(\chi^- \chi^+-1) \log (\chi^-+1)-\log (\chi^-) \log (\chi^++1) \log
   (\chi^-+1)+ \log (\chi^+) \log (\chi^++1) \log (\chi^-+1)+ \\
   & & +2 \log (\chi^-) \log (\chi^- \chi^+-1) \log (\chi^-+1)+2 \log (\chi^++1) \log (\chi^- \chi^+-1) \log (\chi^-+1)+ \\
   & &+\text{Li}_2(\chi^- \chi^+) \log (\chi^-+1)+ \frac{1}{6} \log ^3(\chi^++1)+\frac{1}{3} \log
   ^3(\chi^- \chi^+-1)+\frac{1}{2} \log (\chi^-) \log ^2(\chi^++1)-\\
   & & - \log (\chi^+) \log ^2(\chi^++1)-\log (\chi^++1) \log ^2(\chi^- \chi^+-1)+\log ^2(\chi^+) \log (\chi^++1)- \\
   & & - \log (\chi^-) \log (\chi^+) \log (\chi^++1)-\log ^2(\chi^+) \log (\chi^- \chi^+-1)+  2 \log (\chi^+)
   \log (\chi^++1) \log (\chi^- \chi^+-1)-\\
   & & -\log (\chi^++1) \text{Li}_2(-\chi^-)+\log \left(\frac{\chi^+}{\chi^-}+\chi^+\right) \text{Li}_2(-\chi^+)-\log (\chi^+) \text{Li}_2(\chi^- \chi^+)+\log (\chi^++1) \text{Li}_2(\chi^- \chi^+)+\\
   & &+\log (\chi^-)
   \text{Li}_2\left(\frac{1-\chi^- \chi^+}{\chi^++1}\right)-\log (\chi^+) \text{Li}_2\left(\frac{1-\chi^- \chi^+}{\chi^++1}\right)-\text{Li}_3(-\chi^-)+\text{Li}_3(\chi^-+1)-\text{Li}_3(-\chi^+)+\\
   & & +\text{Li}_3(\chi^-
   \chi^+)-\text{Li}_3\left(\frac{(\chi^-+1) \chi^+}{\chi^++1}\right)+\text{Li}_3(\chi^++1)-\text{Li}_3\left(\frac{\chi^- (\chi^++1)}{\chi^-+1}\right)+ 2 \text{Li}_3\left(\frac{1}{1-\chi^- \chi^+}\right)-\\
   & &-2
   \text{Li}_3\left(\frac{\chi^-+1}{1-\chi^- \chi^+}\right)-2 \text{Li}_3\left(\frac{\chi^++1}{1-\chi^- \chi^+}\right)
   \end{eqnarray*}
}

\end{document}